\documentclass[lettersize,journal]{IEEEtran}
\usepackage{amsmath,amsfonts}
\usepackage{algorithmic}
\usepackage{algorithm}
\usepackage{array}
\usepackage[caption=false,font=normalsize,labelfont=sf,textfont=sf]{subfig}
\usepackage{textcomp}
\usepackage{stfloats}
\usepackage{url}
\usepackage{verbatim}
\usepackage{graphicx}
\usepackage{cite}
\usepackage{multirow}
\begin{document}

\title{CiUAV: A Multi-Task 3D Indoor Localization System for UAVs based on Channel State Information}

\author{Cunyi Yin\dag, Chenwei Wang\dag, Jing Chen, Hao Jiang, Xiren Miao, Shaocong Zheng\\Zhenghua Chen~\IEEEmembership{Senior Member,~IEEE},  
and Hong Yan~\IEEEmembership{Fellow,~IEEE}
\thanks{This work was supported by the Hong Kong Innovation Technology Commission (InnoHK Project CIMDA) and Natural Science Foundation of Fujian Province, China, grant number 2022J01566. (Corresponding author: Jing Chen (chenj@fzu.edu.cn).)

C. Yin, C. Wang and H. Yan are with the CityU-Oxford joint center for intelligent multidimensional data analysis, City University of Hong Kong, Hong Kong (cunyiyin1125@gmail.com; dbw181101@163.com; h.yan@cityu.edu.hk), J. Chen, H. Jiang, X. Miao and S. Zheng are with the College of Electrical Engineering and Automation, Fuzhou University, Fuzhou, China (chenj@fzu.edu.cn; jiangh@fzu.edu.cn; mxr@fzu.edu.cn; 210127054@fzu.edu.cn), Z. Chen is with the Institute for Infocomm Research, Agency for Science, Technology and Research, Singapore (chen0832@e.ntu.edu.sg)

\dag Equal contribution.
}
}

\markboth{Journal of \LaTeX\ Class Files,~Vol.~14, No.~8, August~2021}%
{Shell \MakeLowercase{\textit{et al.}}: A Sample Article Using IEEEtran.cls for IEEE Journals}


\maketitle

\begin{abstract}
Accurate indoor positioning for unmanned aerial vehicles (UAVs) is critical for logistics, surveillance, and emergency response applications, particularly in GPS-denied environments. Existing indoor localization methods, including optical tracking, ultra-wideband, and Bluetooth-based systems, face cost, accuracy, and robustness trade-offs, limiting their practicality for UAV navigation. This paper proposes CiUAV, a novel 3D indoor localization system designed for UAVs, leveraging channel state information (CSI) obtained from low-cost ESP32 IoT-based sensors. The system incorporates a dynamic automatic gain control (AGC) compensation algorithm to mitigate noise and stabilize CSI signals, significantly enhancing the robustness of the measurement. 
Additionally, a multi-task 3D localization model, Sensor-in-Sample (SiS), is introduced to enhance system robustness by addressing challenges related to incomplete sensor data and limited training samples. SiS achieves this by joint training with varying sensor configurations and sample sizes, ensuring reliable performance even in resource-constrained scenarios. 
Experiment results demonstrate that CiUAV achieves a LMSE localization error of 0.2629 m in a 3D space, achieving good accuracy and robustness. The proposed system provides a cost-effective and scalable solution, demonstrating its usefulness for UAV applications in resource-constrained indoor environments.

\end{abstract}

\begin{IEEEkeywords}
Indoor UAVs, channel state information, indoor localization, deep learning.
\end{IEEEkeywords}

\section{Introduction}
\IEEEPARstart{U}{nmanned} Aerial Vehicles (UAVs) have become indispensable for various applications, including logistics, surveillance, and emergency response. Accurate navigation in indoor environments, where GPS signals are unavailable or unreliable, is crucial to unlocking the full potential of UAVs \cite{ali2023comprehensive, kassas2022not}. Precise indoor positioning ensures safe navigation and enables advanced applications, such as autonomous inspections and real-time surveillance, that require centimeter-level accuracy \cite{yin2023overview, qi2024current}. However, achieving reliable and precise positioning in complex indoor environments remains a significant challenge.

Significant advancements are made in sensor technology, with numerous sensors employed in detection and localization tasks \cite{yin2021device, yin2023human}. Existing indoor localization technologies, such as optical tracking systems, Ultra-Wide Band (UWB), and Bluetooth-based methods, are significantly advanced \cite{yin2023overview, hayward2022survey, zou2017winips, yin2021high}. Optical systems offer high accuracy but are expensive and constrained by line-of-sight requirements. UWB systems also provide high accuracy but can be hindered by hardware complexity and deployment challenges. Bluetooth-based positioning is cost-effective but lacks the precision necessary for UAV navigation. These limitations underscore the need for alternatives that balance cost, accuracy, and robustness. Recent advances in wireless communications have introduced channel state information (CSI) as a promising solution for indoor localization \cite{sohrabi2022learning}. CSI extracted from WiFi captures fine-grained details about the signal propagation environment. CSI-based localization becomes a viable solution for indoor UAV localization. However, most existing research has focused on 2D localization and has yet to address the challenges of achieving 3D UAV localization using CSI.

Despite the significant potential of CSI, several challenges remain in applying it to precise indoor UAV localization. Firstly, no 3D localization method is designed explicitly for CSI, particularly for UAV applications, creating a gap in research and practical implementation. Also, indoor UAVs impose stringent positioning accuracy and stability requirements due to their dynamic characteristics. However, the amplitude and phase components of CSI are highly susceptible to noise and environmental factors, which can significantly degrade performance \cite{huang2024hybrid}. Moreover, CSI-based localization methods often assume the availability of complete data from all sensors and extensive training datasets to achieve high accuracy. However, in practical applications, network and communication disruptions may result in incomplete sensor data or limited samples, posing significant challenges to system performance.
 
To address these issues, we developed CiUAV, a low-cost system utilizing CSI and the ESP32-S3 IoT chip. CiUAV captures CSI data from UAVs without additional onboard equipment, offering a device-free solution ideal for payload-limited drones. A dynamic AGC compensation algorithm is introduced to mitigate indoor CSI signal distortions by converting gain values into linear scaling factors, enhancing system robustness. Additionally, we propose a multi-task joint 3D localization model that leverages collaborative cross-task learning to adapt to varying sensor configurations and limited training data availability. The model ensures robust and accurate localization by sharing a feature extraction layer across tasks, even under incomplete sensor input or reduced data scenarios. Experiment results demonstrate the reliability, scalability, and practicality of the system for resource-limited UAV applications.

We summarize the contributions of this work as follows:
\begin{itemize}
\item{We design CiUAV, a novel CSI-based 3D spatial localization method designed specifically for indoor UAV environments. A low-cost indoor UAV localization system is developed using the ESP32 IoT chip, which passively captures CSI signals from UAVs without requiring additional onboard equipment, providing a device-free solution for UAV localization.}
\item{A dynamic AGC compensation (DAC) algorithm based on an analysis of UAV CSI noise mechanisms is leveraged to reduce the impact of noise on CSI measurements. The DAC is further enhanced by integrating a CSI outlier processing method, resulting in an optimized CSI representation with more distinct features.}
\item{We propose Sensor-in-Sample (SiS), a multi-task 3D localization model that optimizes sensor quantity and training data size in CSI tasks. The model effectively mitigates task conflicts, improves generalization, and extracts critical CSI features, enabling accurate localization and reliable performance despite sensor or data limitations by sharing a feature extraction layer across tasks.}
\end{itemize}

\section{Related Work}
\subsection{WiFi-Based Indoor Positioning}
WiFi received signal strength indicator (RSSI) has been widely studied for indoor positioning due to its availability in common devices. Geometric methods use path loss models with trilateration but suffer from environmental interference, while fingerprinting, enhanced by machine learning \cite{rezgui2017efficient, li2015feature}, improves performance only moderately. Deep learning approaches such as RNN-based methods \cite{qian2021supervised} and CNNLoc \cite{song2019novel} enhance feature extraction, but do not meet the precision needed for UAV localization.

CSI, which offers subcarrier-level frequency domain data for finer signal propagation analysis, outperforms RSSI. CSI-based methods include distance estimation, array antenna-based techniques, and machine learning fingerprinting. Distance estimation mitigates multipath effects for improved trilateration accuracy, as shown in FILA \cite{wu2012fila} and LiFS \cite{wang2016lifs}. ArrayTrack \cite{xiong2013arraytrack} and SpotFi \cite{kotaru2015spotfi} enhance precision via AoA estimation but require specialized hardware. Machine learning fingerprinting systems such as DeepFi \cite{wang2015deepfi} and CiFi \cite{wang2017cifi, wang2017biloc, choi2017deep} leverage deep learning to extract high-level CSI features, achieving superior 2D positioning. However, most methods focus on 2D scenarios, falling short in dynamic 3D environments needed for UAVs.
The key challenges for CSI-based UAV localization are managing multipath propagation, achieving high-precision 3D resolution, and ensuring robustness in dynamic settings. Advanced modeling techniques leveraging CSI's multidimensional nature are crucial for accurate 3D localization.

\subsection{CSI-Based Passive Sniffing}
Indoor micro UAVs commonly use WiFi to transmit video, images, and sensor data, enabling operators to monitor and control flights \cite{bisio2024rf}. The prevalence of WiFi in indoor spaces facilitates CSI collection from UAVs, but tools must be universal and non-invasive, avoiding hardware modifications or interference \cite{li2023fusing}.
Passive sniffing addresses these needs by collecting CSI without requiring UAV communication links, relying on their normal operations \cite{yin2024powerskel}. Using ICMP echo requests to extract CSI from reply packets, this method ensures efficient, non-intrusive data acquisition \cite{miranda2021estimating}.

Traditional tools like Intel IWL5300 \cite{suthar2021multiclass} and Atheros WiFi cards rely on outdated configurations, while Nexmon firmware \cite{dahal2024comparison} requires specific WiFi chips, limiting their practicality. These challenges underline the need for universal, non-invasive passive sniffing tools for UAVs.
Applying CSI to UAV localization faces additional challenges in dynamic environments. Traditional passive methods, designed for static devices, struggle with multipath propagation, frequency offsets, and 3D spatial analysis. Real-time localization for UAVs' rapid movements further increases complexity. Advancing indoor UAV positioning requires methods that address these challenges while leveraging CSI for precise, efficient 3D localization.

\subsection{Multi-Task Learning}
Multi-task learning (MTL) leverages shared features across related tasks to improve generalization and performance in fields such as NLP and computer vision \cite{li2021interactive}. Early MTL methods used hard parameter sharing, where tasks share a common feature extraction layer, or soft parameter sharing, which links independent models through task-specific constraints \cite{vandenhende2021multi}. These approaches mitigate overfitting and improve information utilization. Recently, MTL has evolved to incorporate dynamic mechanisms, adjusting task processing using contextual information to capture inter-task relationships better and enhance robustness in complex environments \cite{wang2024skeleton}.

In CSI-based indoor UAV localization, traditional single-task models focus solely on position prediction, overlooking multi-dimensional factors like signal propagation and noise. Intertask competition in MTL can cause bottlenecks, especially in correlated or complex environments. A dynamic MTL framework that co-optimizes multiple tasks and efficiently utilizes CSI's multi-dimensional characteristics is essential. The framework improves adaptability, robustness, and accuracy in complex UAV localization scenarios.

\section{Methodology}
\subsection{CSI-Based Indoor UAV System Design}
\begin{figure*}[!t]
\centering
\includegraphics[width=6.5in]{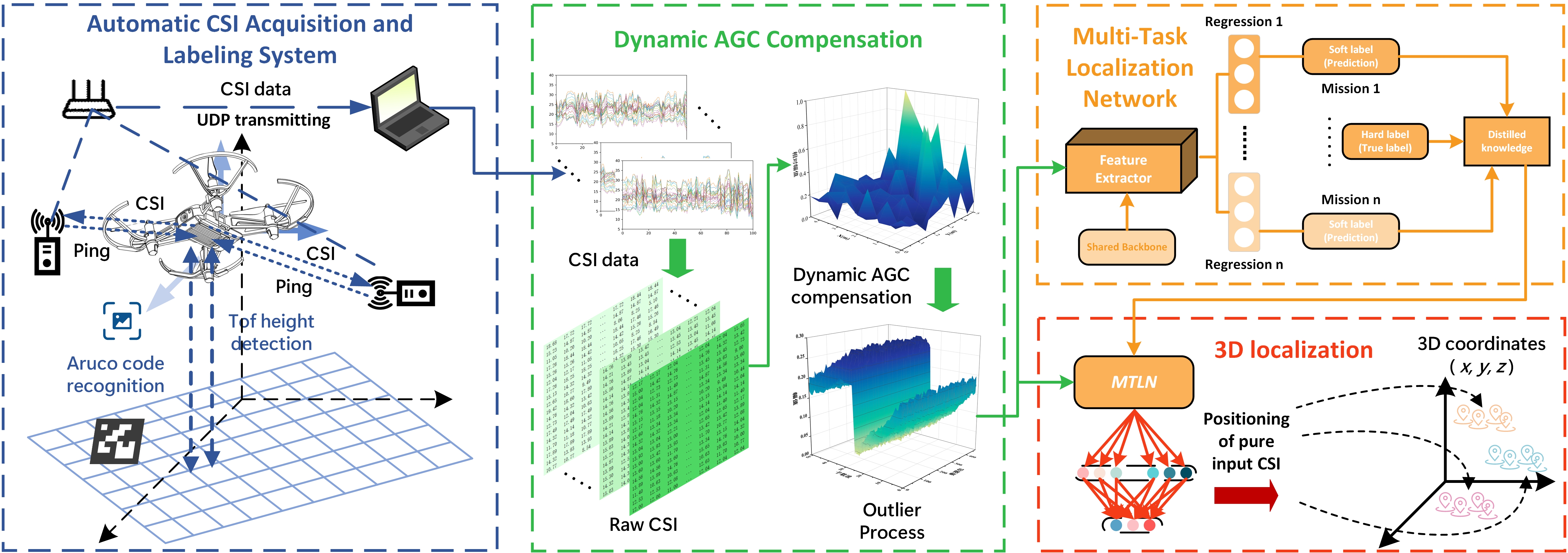}%
\caption{Overview of the CiUAV framework, comprising data acquisition, dynamic AGC compensation, model training, and 3D localization. CSI sensors capture signals from the UAV, which uses an onboard camera and ToF sensor for ground-truth 3D positions. A dynamic AGC algorithm corrects amplitude distortions, while the SiS network predicts 3D coordinates with minimal data and sensor requirements, enabling real-time localization.}
\label{fig_1}
\end{figure*}

The CiUAV framework comprises four primary components: data acquisition, dynamic AGC compensation, model training, and 3D localization, as illustrated in Fig. \ref{fig_1}.
During the data acquisition phase, multiple custom-designed CSI sensors are deployed within the designated area to capture CSI signals from the UAV in flight passively. During the flight, the UAV utilizes its onboard camera to detect Aruco markers to obtain 2D planar coordinates and a Time-of-Flight (ToF) sensor to measure altitude. The CiUAV can obtain the UAV's 3D position, which serves as the ground-truth label. The CSI data and the UAV's 3D coordinate information are transmitted to the host computer via wireless communication.

We propose a dynamic AGC compensation algorithm to address amplitude distortions caused by AGC circuits in CSI sensors. This algorithm retrieves real-time hardware-level AGC gain values, converts them into linear scaling factors, and applies these factors to compensate for CSI amplitudes. By eliminating AGC-induced distortions, the algorithm accurately restores the signals' actual amplitude characteristics, ensuring their suitability for high-precision indoor UAV positioning tasks.
For model training, we developed a multi-task learning-based network, termed Sensor-in-Sample (SiS), to predict the UAV's 3D coordinates from CSI data. The SiS network is designed to minimize the number of required training instances and CSI sensors by leveraging task generalization and knowledge sharing during joint training. The SiS enables accurate prediction of the UAV's 3D location based on CSI data.
Once the SiS network is trained offline, the framework can deliver precise real-time 3D position estimation for UAVs using only raw CSI input, regardless of variations in sensor availability or data completeness.

\subsection{Data Acquisition}
A CSI data acquisition and labeling system was developed by integrating vision-based Aruco markers with a ToF sensor. Custom firmware is created to construct CSI sensors using the ESP32-S3 microcontroller, enabling CSI collection by UAVs. These sensors utilize ICMP to send ICMP Echo Request packets (Ping) to the UAV, with the CSI information parsed from the ICMP Echo Reply packets returned by the UAV. This passive sniffing process facilitates localization without requiring the UAV to carry additional hardware or establish an active communication link. This non-intrusive approach is UAV-friendly and requires no modifications to the UAV itself.

Multiple CSI sensors are deployed within the environment to passively sniff the UAV's signals and collect real-time CSI data. Using the UDP protocol, these sensors support WiFi communication and transmit the collected CSI data to the host computer in real-time.
The UAV begins at an initial takeoff point and uses its onboard camera to detect and identify Aruco markers within its visual range. Simultaneously, the UAV transmits captured video data and ToF sensor height measurements to the host computer via WiFi. The UAV's 2D planar coordinates are derived from the decoded Aruco markers, while the ToF sensor provides altitude information. These measurements form the UAV's 3D coordinate data, serving as label information for the corresponding CSI data.

The host computer uses these coordinates to control the UAV's movement through a Proportional-Integral-Derivative algorithm. Upon reaching a designated data collection point, the UAV's camera is calibrated and aligned, after which it sends a command to the CSI sensors to acquire CSI data. The collected CSI data and the corresponding 3D coordinates are saved in real-time as labeled training data. The host employs a shortest path planning algorithm to navigate the UAV to the nearest data collection point efficiently. This iterative process repeats until the data collection is complete.

\subsection{Dynamic AGC Compensation Algorithm}
The AGC circuit dynamically adjusts the signal gain to ensure the stability of WiFi communication. However, this process introduces amplitude distortion in the CSI, masking variations in signal amplitude, and making it difficult for CSI values to reflect positional differences directly. Consequently, AGC-related distortion poses significant challenges for CSI-based indoor localization. To improve the reliability of localization based on CSI amplitude, a dynamic AGC compensation algorithm is proposed that leverages real-time AGC gain information to adjust CSI data while preserving the standard communication functionality of the devices.

The ESP32-S3 chip, which is the core of the CSI sensors in this paper, offers real-time access to AGC gain values through its latest firmware development interface. The feature enables the direct retrieval of AGC gain information without relying on indirect inference or estimation. By reading the real-time AGC gain values from the hardware layer, the proposed algorithm better understands the actual signal gain adjustments, enabling precise and stable compensation.

In the CiUAV system, a fixed-power signal is transmitted from the sender and propagates through space to the receiver. At the receiver, the signal is amplified based on its strength to maintain communication stability, a process controlled by the AGC circuit. This amplification introduces errors in the processed CSI data, as the AGC circuit affects the extracted CSI amplitude fingerprints, causing them to deviate from the expected path loss model. These errors significantly degrade localization accuracy. The amplitude error introduced by AGC can be expressed as:
\begin{equation}
H(k)=\alpha_{\mathrm{AGC}}|H(k)| \mathrm{e}^{\mathrm{j} \angle H(k)},
\end{equation}
where $\alpha_{\mathrm{AGC}}$ represents the amplitude error factor, and $\mathrm{AGC}$ is the gain.
The proposed algorithm uses the AGC gain values directly to perform compensation. Specifically, the AGC gain values are converted from decibels to linear scale, and the reciprocal value of this value is utilized as a scaling factor for the CSI amplitude compensation. The scaling factor is calculated as follows:
\begin{equation}
\rho=\frac{1}{10^{A G C _{\operatorname{gain} / 20}}},
\end{equation}
where $\rho$ represents the scaling factor and $A G C _{\operatorname{gain}}$ denotes the AGC gain in decibels.
The scaling factor is applied linearly to each CSI subcarrier to compensate for amplitude distortion, as expressed by the following equation:
\begin{equation}
\tilde{H}(k)=\rho^* H(k)=\frac{H(k)}{10^{A G C \operatorname{gain}_k / 20}},
\end{equation}
where $\tilde{H}(k)$ represents the compensated CSI, and it includes the real and imaginary components of the original CSI data. Both components are corrected simultaneously, and the compensated CSI amplitude is represented as:
\begin{equation}
|\tilde{H}(\mathrm{k})|=\sqrt{\left(\tilde{H}_{\text {real }}(k)\right)^2+\left(\tilde{H}_{\text {imag }}(k)\right)^2} H(k),
\end{equation}
where $\tilde{H}_{\text {real }}(k)$ and $\tilde{H}_{\text {imag }}(k)$ are the real and imaginary parts of the CSI after linear compensation.
The above steps restore the actual amplitude characteristics of the CSI captured by the sensors. 

Given the complexity of indoor communication environments, significant disturbances can affect CSI data, affecting high-precision UAV positioning. Based on real-time AGC gain feedback, the proposed compensation strategy effectively tracks and mitigates the impact of AGC-induced distortion. By dynamically eliminating AGC-related effects, this method significantly improves the accuracy of indoor UAV localization systems based on CSI.

\subsection{SiS: Multi-task 3D Localization Model}
Traditional CSI-based indoor positioning methods rely on extensive training samples and comprehensive sensor deployments to achieve high spatial resolution and positioning accuracy. However, these approaches face hardware constraints, significant data collection overhead, and poor generalization when limited or unevenly distributed training data.

To address these issues, this paper proposes a multi-task joint 3D localization model, SiS, which focuses on optimizing sensor configurations and training data usage. The model promotes knowledge sharing and enhances generalization across tasks by incorporating cross-task joint training, ensuring reliable performance under practical constraints.

The proposed framework contains the feature extraction process, task-specific regression, and integration of regularization components to improve the sensor and sample efficiency. This structured design ensures that the model not only achieves high positioning accuracy but also effectively optimizes resource utilization.
The input data consists of CSI amplitude-only signals represented as $\mathbf{X} \in \mathbb{R}^{N \times S \times f}$, where $N$ denotes the number of samples, $S$ the number of sensors, and $f$ the number of subcarrier amplitude values per sensor. The corresponding labels are the 3D coordinates of the drone, expressed as $\mathbf{Y} \in \mathbb{R}^{N \times S \times D}$, with $D$ = 3 representing the dimensions $x$, $y$, and $z$.

The model first employs a feature extractor $F_{e}$ to derive feature representations from the input data:
\begin{equation}
\mathbf{H}=F_{\text {e}}(\mathbf{X}),
\end{equation}
where $\mathbf{H} \in \mathbb{R}^{N \times S \times f_h}$, $f_h$ represents the dimension of the extracted features. 
The feature extractor is designed to identify meaningful patterns and structural information within the input data, providing a robust representation for downstream tasks.

The extracted feature representations $\mathbf{H}$ are then mapped to the 3D coordinate space through a regression module $F_{r}$, resulting in the predicted positions for each sensor:
\begin{equation}
\hat{\mathbf{Y}}=F_{\text {r}}(\mathbf{H}),
\end{equation}
where $\hat{\mathbf{Y}} \in \mathbb{R}^{N \times S \times D}$. The loss function of the model is designed to achieve the dual objectives of reducing the number of training samples and CSI sensors while maintaining high positioning accuracy. 
The prediction error loss is defined as:
\begin{equation}
\mathcal{L}_{\text {pred }}=\frac{1}{N \times S} \sum_{i=1}^N \sum_{s=1}^S\left\|\hat{\mathbf{Y}}_{i, s,:}-\mathbf{Y}_{i, s,:}\right\|_2^2,
\end{equation}
where $\hat{\mathbf{Y}}_{i, s,:}$ and $\mathbf{Y}_{i, s,:}$ denote the predicted and ground truth coordinates for the $i$-th sample and the $s$-th sensor, respectively. 
To minimize the number of sensors, a sparsity regularization term is introduced to constrain sensor usage:
\begin{equation}
\mathcal{L}_{\text {sor}}=\lambda_s\left\|\mathbf{w}_s\right\|_1,
\end{equation}
where $\mathbf{w}_s$ represents the sensor weight vector, and $\lambda_s$ is the sparsity regularization coefficient. This term identifies and retains the most informative sensors while reducing redundant deployments.
In addition, a sample selection regularization term is included to decrease dependence on large-scale training data. 
By introducing a sample weight vector $\mathbf{v}$, the model emphasizes samples with higher contributions to the positioning task:
\begin{equation}
\mathcal{L}_{\text {sam}}=\frac{1}{S} \sum_{s=1}^S\left\|\mathbf{v} \odot\left(\hat{\mathbf{Y}}_{:, s,:}-\mathbf{Y}_{:, s,:}\right)\right\|_2^2,
\end{equation}
where $\mathbf{v} \in \mathbb{R}^N$ is the sample weight vector, and $\odot$ denotes element-wise multiplication. Optimizing $\mathbf{v}$ allows the model to prioritize high-value samples, thus reducing the reliance on extensive datasets.

The total loss function is expressed as:
\begin{equation}
\mathcal{L}_{\text {total }}=\mathcal{L}_{\text {pred }}+\mathcal{L}_{\text {sor }}+\lambda_v \mathcal{L}_{\text {sam }},
\end{equation}
where $\lambda_v$ is the regularization coefficient for sample selection. 
This combined loss ensures a balance between positioning accuracy, sensor sparsity, and sample efficiency, enabling the model to perform effectively under resource constraints.

Through this design, the model effectively learns the relationships between sensor configurations, training data size, and position information, enabling precise 3D drone localization without relying on extensive sensor deployments or large-scale training data. By leveraging joint multi-task training, the model resolves task conflicts and enhances generalization across tasks, providing a flexible and robust solution for UAV localization in complex indoor environments.

\section{Experiments}
\subsection{Setup}
\begin{figure}[!t]
\centering
\includegraphics[width=3.4in]{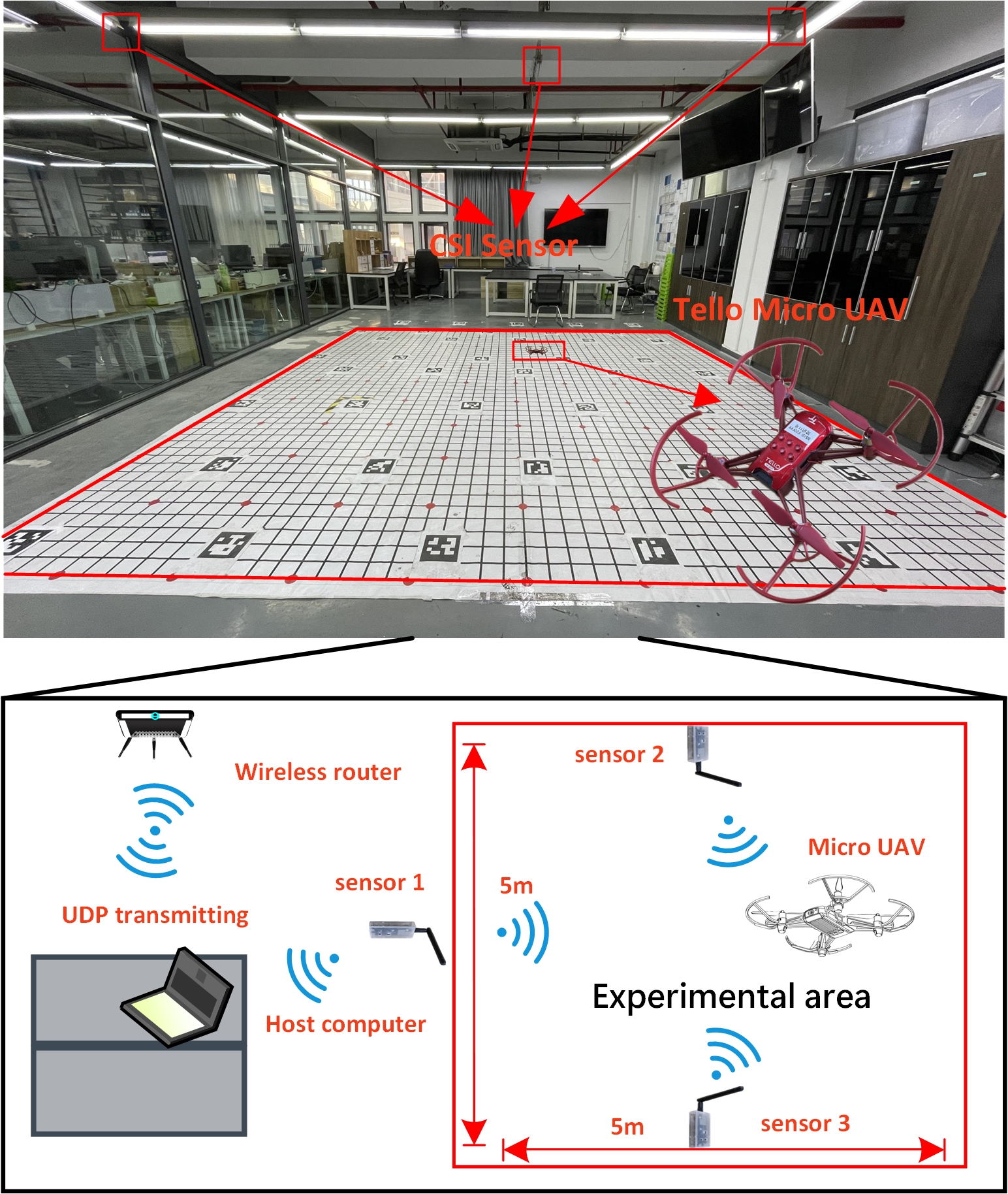}%
\caption{The layout of the experiment environment and Planar graph. Passive sniffing of UAVs by CSI sensors.}
\label{fig_2}
\end{figure}

The experiment setup is illustrated in Fig. \ref{fig_2}, showing the overall experiment environment and its layout. The experiment was carried out in a typical indoor setting, where three CSI sensors were deployed in a triangular configuration at a height of 2.5 m on the ceiling. The UAV operated within a 5 m $\times$ 5 m carpeted area designated for flight and data collection. A TL-XDR3010 wireless router was used to establish the local area network communication link. The three CSI sensors were connected to the router via a one-touch network configuration, while the indoor UAV was also connected to the router through WiFi.
A host computer (Intel Core i7-10700 CPU, 64 GB RAM, NVIDIA GeForce RTX 4070 Ti SUPER) was used for data storage and computation, interfacing with the router to maintain network connectivity. All experiment devices operated within the same LAN, allowing seamless communication through the router. The wireless network operated in a 2.4 GHz frequency band, with data transmission utilizing the UDP protocol. UAV control and data streams were transmitted through the Tello UAV custom API interface.

The experiment data were automatically collected using the CSI acquisition and labeling system described in Section II, Part A. Specifically, the CSI sensors were configured with a data sampling rate of 50 Hz. The host computer controlled the UAV's movement within the data collection area, guiding it to 121 predefined red grid points at heights between 0.6 m and 2.0 m. At each point and height of the grid, 500 frames of CSI data were recorded.
We randomly chose three time periods to collect data to ensure generalization of the data and used them as training sets. A separate dataset was collected exclusively for testing purposes. The training set comprised 77,000 samples, while the test set contained 33,000.

The proposed multi-task network for indoor drone localization is configured with specific parameters to optimize its performance and facilitate efficient resource utilization. The input CSI data consists of $N$=10,000 samples, each containing $S$=3 sensors with $f$=50 subcarrier amplitude values. The extracted features have a dimensionality of $F_h$=128, determined by the feature extractor $F_e$. 

The regularization terms are weighted with $\lambda_s$=0.01 for sensor sparsity and $\lambda_v$=0.1 for sample selection, balancing the objectives of minimizing the sensor and training sample requirements. The model is trained with Adam Optimizer with an initial learning rate of 0.0001, a batch size of 32, and a total of 400 epochs. All computations are conducted using the PyTorch framework. These configurations are designed to ensure that the model achieves accurate 3D positioning while effectively addressing limited sensors and training data constraints.
The Mean Absolute Error ($MAE$), the Localization Mean Squared Error ($LMSE$), and the Coefficient of Determination ($R^2$) of the positioning error are utilized to evaluate the effect of the positioning model, where $R^2$ closer to 1 means better localization.

\subsection{Result}
\begin{figure}[!t]
\centering
\includegraphics[width=3.4in]{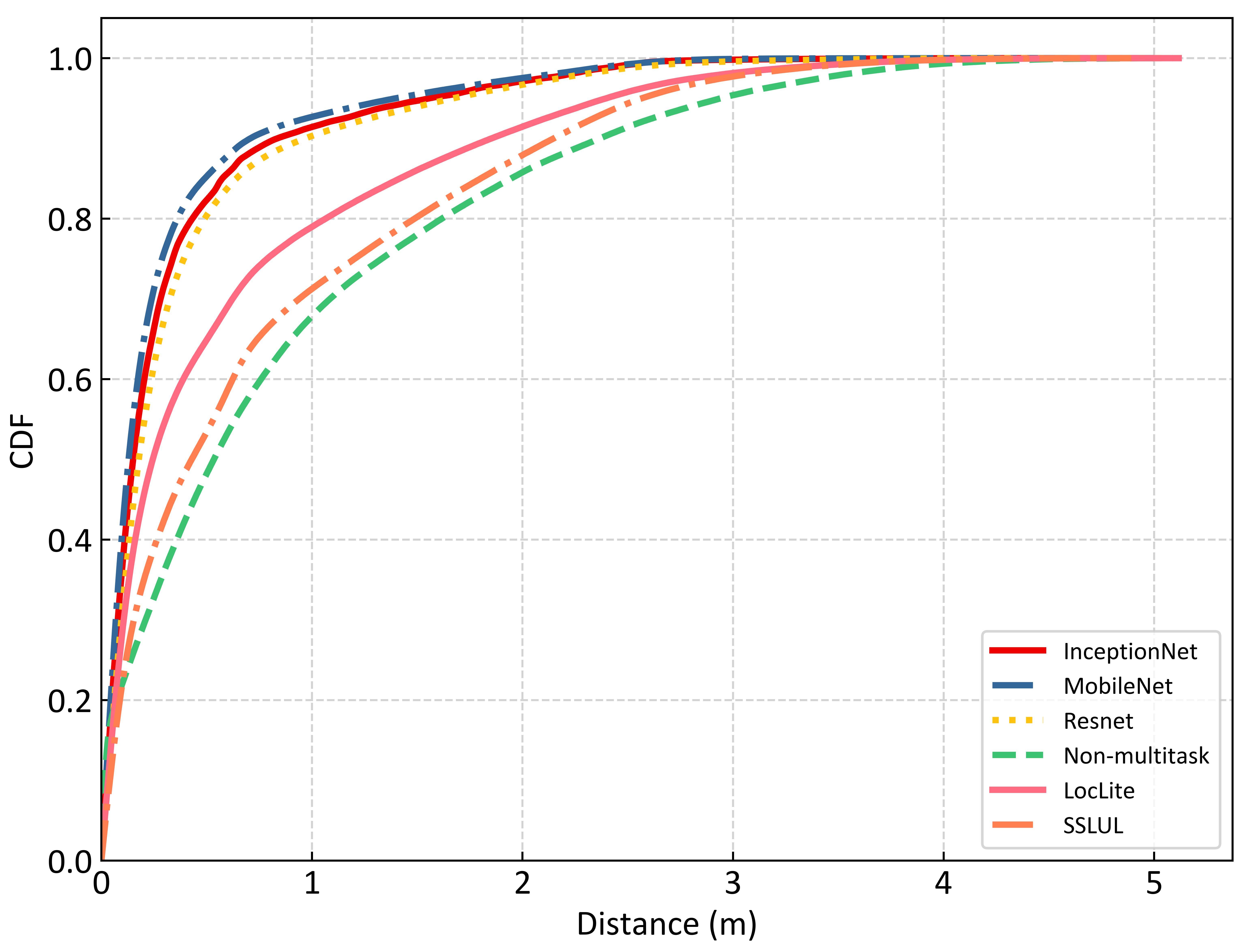}%
\caption{CDF curves of localization errors for the multi-task 3D localization framework using InceptionNet, MobileNet, ResNet, and a baseline single-task model. The multi-task models outperform the baseline, with MobileNet achieving the best accuracy and robustness.}
\label{fig_3}
\end{figure}

Fig. \ref{fig_3} presents the cumulative distribution function (CDF) curves of localization errors for the proposed multi-task 3D localization framework using three feature extractors (InceptionNet, MobileNet, and ResNet), as well as baseline single-task model, Loclite \cite{vuckovic2024csi} and SSLUL \cite{dash2024self}. The LocLite and SSLUL are state-of-the-art (sota) localization methods based on CSI and deep learning. The results highlight the substantial performance gains achieved by the multi-task framework.
The multi-task framework significantly outperforms the non-multi-task model, as indicated by the steep CDF curves of all three feature extractors. The InceptionNet achieves an $LMSE$ of 0.3684 m, demonstrating intense localization precision. The MobileNet, with an $LMSE$ of 0.2629 m, offers the best overall accuracy, as reflected by its superior CDF curve, where more than 80\% of errors are within 0.25 m. The ResNet, achieving an $LMSE$ of 0.3132 m, balances precision and robustness, effectively reducing high-error outliers. In contrast, the single-task baseline model has an $LMSE$ of 1.6947 m, with a significantly flatter CDF curve, indicating much poorer performance and less reliable predictions.

The CDF curves further illustrate the robustness of the multi-task framework under different conditions. The multi-task models achieve 80\% of localization errors below 0.5 m. In contrast, the Non-multitask model struggles to achieve the same level of precision, with a much more significant proportion of errors exceeding 0.5 m. The results of the multi-task framework also significantly outperform the sota CSI-based LocLite and SSLUL.
We chose InceptionNet, MobileNet, and ResNet because these models are computationally efficient and well-suited to the characteristics of CSI data, which have limited dimensions. Employing overly complex models on such data would likely result in overfitting, ultimately degrading localization performance. The experiment results validate this decision, as all three feature extractors achieve high accuracy and robust performance when integrated into the multi-task framework.
The proposed multi-task 3D localization framework significantly improves accuracy and robustness over the baseline, effectively leveraging shared feature representations to enhance model generalization. These results highlight the framework's potential as a scalable and efficient solution for UAV localization, particularly in resource-constrained or dynamic environments.

\begin{table}[b]
\caption{Ablation study results for different data processing configurations}
\centering
\begin{tabular}{|cc|ccc|}
\hline
\multicolumn{2}{|c|}{Data process}                  & \multicolumn{3}{c|}{Performance}                                   \\ \hline
\multicolumn{1}{|c|}{Dynamic AGC (DAC)} & Hample+outelier & \multicolumn{1}{c|}{$MAE$}    & \multicolumn{1}{c|}{$LMSE$}   & $R^2$      \\ \hline
\multicolumn{1}{|c|}{}            &                 & \multicolumn{1}{c|}{0.5875} & \multicolumn{1}{c|}{0.3451} & 0.8112 \\ \hline
\multicolumn{1}{|c|}{$\surd$}           &                 & \multicolumn{1}{c|}{0.5366} & \multicolumn{1}{c|}{0.2879} & 0.8419 \\ \hline
\multicolumn{1}{|c|}{}            & $\surd$               & \multicolumn{1}{c|}{0.5596} & \multicolumn{1}{c|}{0.3132} & 0.8283 \\ \hline
\multicolumn{1}{|c|}{$\surd$}           & $\surd$               & \multicolumn{1}{c|}{0.5127} & \multicolumn{1}{c|}{0.2629} & 0.8559 \\ \hline
\end{tabular}
\end{table}
\subsection{Ablation Study}
We conducted ablation experiments under different data processing configurations to assess the impact of DAC and Hample data processing modules with outliers on localization performance. 
Table I presents the variations in performance metrics, including $MAE$, $LMSE$, and $R^2$, with and without these modules.
The results reveal that the incorporation of DAC significantly improves the accuracy of localization. Without DAC, $ MAE$ and $ LMSE$ increase more than 5 cm and 5.7 cm, respectively, while $R^2$ decreases by 3 cm. The result underscores the adverse impact of amplitude distortion caused by the AGC circuit on localization accuracy. By utilizing a dynamic compensation mechanism, the DAC effectively mitigates this distortion, improving the stability and precision of the localization system.

Additional improvements are observed when Hample filtering and outlier handling are applied to the CSI processed by DAC. Specifically, compared to the use of DAC alone, the Hample with outlier module reduces $MAE$ and $LMSE$ by 2.7 cm and 3.2 cm, respectively, and improves $R^2$ by 1.7 cm. The result demonstrates the effectiveness of Hample filtering and outlier handling in suppressing noise and removing anomalous data points within the CSI, further enhancing the model's robustness.
Combining DAC and Hample with outlier modules is essential to improve indoor drone localization performance. DAC addresses amplitude distortion at its source, reducing the impact of artifacts introduced by the AGC circuit. At the same time, the Hample with an outlier module optimizes signal quality by eliminating noise and anomalies. Together, these modules improve the quality of CSI signals, providing more reliable inputs for subsequent feature extraction and localization tasks, and significantly improving localization accuracy.
\begin{figure}[!t]
\centering
\includegraphics[width=3.5in]{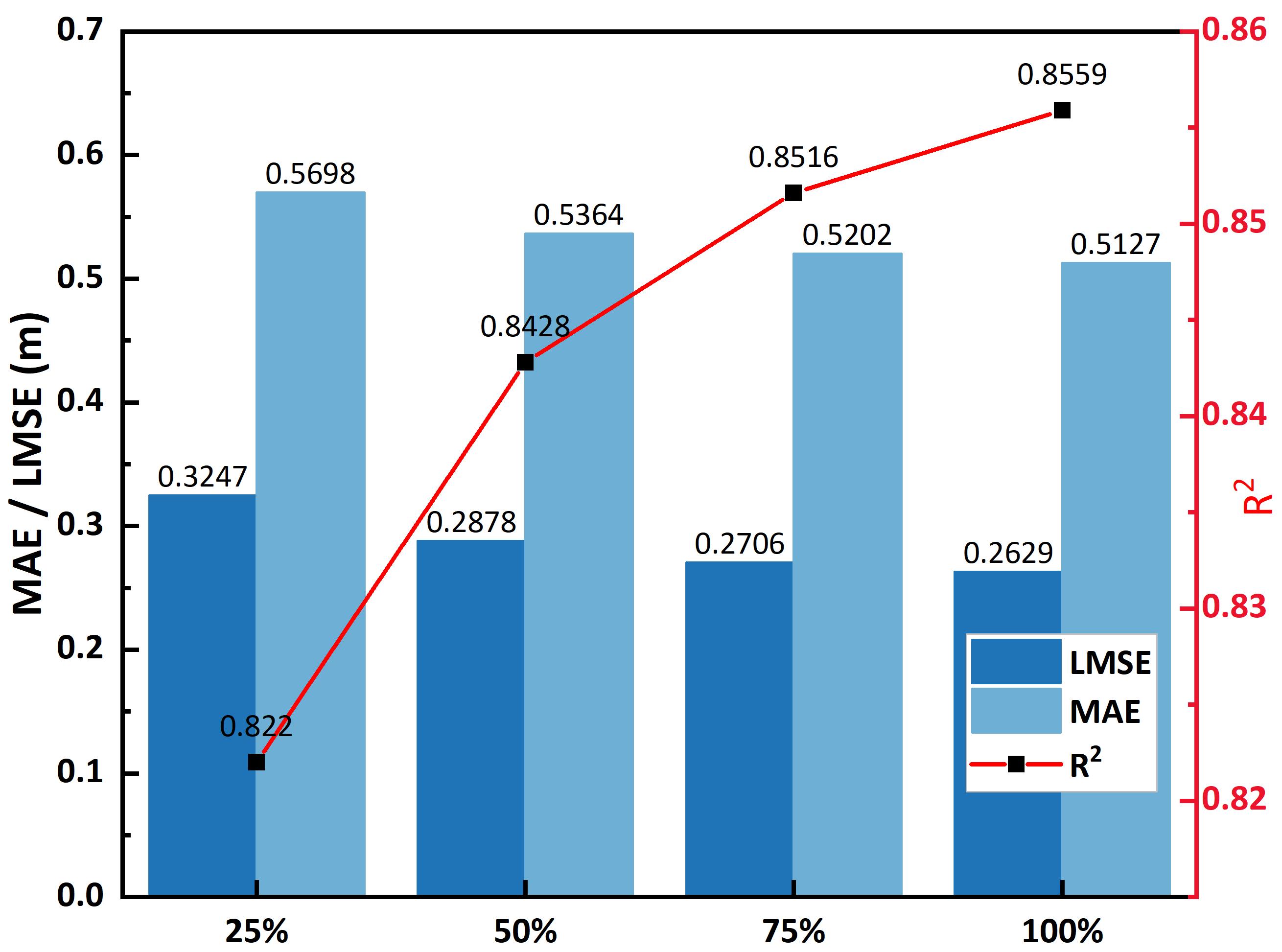}%
\caption{Relationship between training sample proportion and localization performance metrics. The SiS model demonstrates robust adaptability and minimal performance degradation with reduced data.}
\label{fig_4}
\end{figure}

\begin{figure}[!t]
\centering
\includegraphics[width=3.4in]{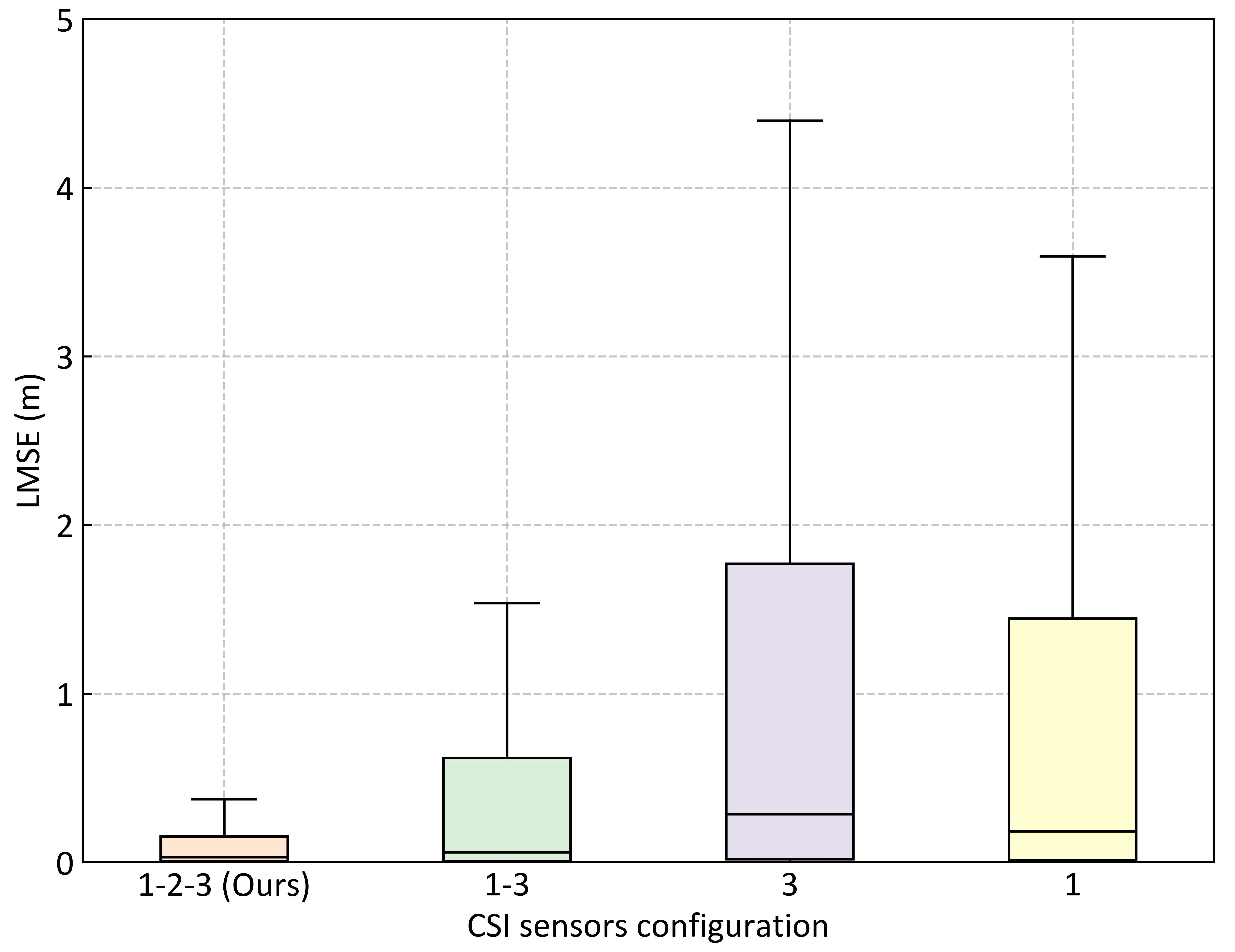}%
\caption{Localization error distributions for different sensor configurations. The ``1-2-3 (Ours)'' configuration achieves the best performance, showcasing the SiS model's ability to utilize complete sensor data effectively. Other configurations highlight the model's adaptability under reduced sensor conditions.}
\label{fig_5}
\end{figure}
\subsection{Analysis of the Sample Size}
To evaluate the adaptability of the proposed SiS model to varying training data sizes, experiments were conducted using training sample proportions of 25\%, 50\%, 75\%, and 100\%. Fig. \ref{fig_4} illustrates the relationship between training sample proportion and evaluation metrics ($MAE$, $LMSE$, and $R^2$). 
The results demonstrate that the model maintains robust localization performance even as the training sample size decreases. For instance, when using only 25\% of the training data, the $MAE$ increases from 0.2629 m (100\%) to 0.3247 m, and the $LMSE$ increases from 0.5127 m to 0.5698 m. Similarly, the $R^2$ value slightly reduces, decreasing from 0.8559 (100\%) to 0.822 (25\%). These results highlight the model's ability to generalize effectively under data-limited conditions. The key to this adaptability lies in the multi-task learning framework, which leverages shared feature representations across tasks. By simultaneously training on varying training data proportions, the model learns to compensate for the reduced availability of data in specific scenarios. This approach ensures that even with limited training samples, the model can adaptively combine knowledge from tasks with richer data, thereby mitigating performance degradation.

Furthermore, the multi-task framework reduces the reliance on large training datasets by promoting data efficiency. For example, when training with only 25\% of the data, the performance gap is minimal compared to the entire dataset scenario, with differences in $MAE$ and $LMSE$ kept under 0.07 m. This demonstrates the ability of multi-task learning to maintain reliable performance even when training data is scarce. The practical implications of this are significant, especially in real-world applications where environmental factors, cost, or time may constrain data collection. By optimizing the utilization of available data through multi-task learning, the SiS model ensures robust and accurate localization performance, offering a scalable and efficient solution for UAV positioning in resource-limited scenarios. The SiS model achieves strong adaptability to varying training data sizes through multi-task learning. By leveraging shared feature representations and cross-task optimization, the model effectively reduces the impact of data limitations, ensuring efficient and reliable 3D localization.

\subsection{Analysis on the Sensor Configuration Optimization}
Experiments were conducted using 4 sensor configurations: ``1-2-3 (Ours)'',  indicates that all configured sensors are used for joint training, i.e., single, 2, and 3 all sensor data. ``1-3'', where data from sensors 1 and 3 are jointly trained. ``3'', with only sensor 3 data trained, and ``1'', with only sensor 1 data trained. These four configurations reflect the diverse setups evaluated to demonstrate the adaptability and robustness of the proposed multi-task learning model. Experiments were conducted using configurations of 1, 2, and 3 sensors to evaluate the adaptability of the proposed SiS model to varying numbers of CSI sensors. Fig. \ref{fig_5} illustrates the localization error distributions for each configuration, and Table II details the metrics $MAE$, $LMSE$, and $R^2$.

\begin{table}[]
\centering
\caption{Localization performance metrics for various sensor configurations}
\label{tab:my-table}
\begin{tabular}{|c|ccc|}
\hline
\multirow{2}{*}{\textbf{CSI Sensors Configuration}} & \multicolumn{3}{c|}{\textbf{Performance}}                                   \\ \cline{2-4} 
                             & \multicolumn{1}{c|}{$\mathbf{MAE}$}    & \multicolumn{1}{c|}{$\mathbf{LMSE}$}   & $\mathbf{R^2}$      \\ \hline
3                       & \multicolumn{1}{c|}{1.3018} & \multicolumn{1}{c|}{1.6947} & 0.0717 \\ \hline
1                       & \multicolumn{1}{c|}{1.1563} & \multicolumn{1}{c|}{1.3371} & 0.2676 \\ \hline
1-3                         & \multicolumn{1}{c|}{0.9968} & \multicolumn{1}{c|}{0.9936} & 0.4560 \\ \hline
1-2-3 (Ours)           & \multicolumn{1}{c|}{0.5127} & \multicolumn{1}{c|}{0.2629} & 0.8559 \\ \hline
\end{tabular}
\end{table}

The results demonstrate that the ``1-2-3 (Ours)'' configuration achieves the best performance, with $MAE$ and $LMSE$ values of 0.5127 m and 0.2629 m, respectively, and an $R^2$ of 0.8559, indicating strong correlation and accuracy in 3D localization. In comparison, the ``1'' configuration (using only sensor 1) increases $MAE$ and $LMSE$ to 1.1563 m and 1.3371 m, respectively, with a significantly reduced $R^2$ of 0.2676. These findings highlight the SiS model's ability to utilize data from all sensors fully, enabling superior localization performance.
The SiS model's multi-task collaborative learning framework is critical to these results. The model ensures robust localization even when the number of sensors is reduced by leveraging shared feature extraction layers and knowledge transfer across tasks. Especially, the ``1-3'' configuration achieves intermediate performance, with $MAE$ and $LMSE$ of 0.9968 m and 0.9936 m, and an $R^2$ of 0.4560, showcasing the model's adaptability under varying sensor conditions.
The ``1-2-3 (Ours)'' configuration demonstrates the optimal use of sensor data and validates the effectiveness of the proposed multi-task learning approach. The SiS model achieves reliable generalization and precise localization performance by jointly training on multiple sensor configurations. This ability to maximize accuracy with all available sensors while maintaining adaptability under reduced sensor conditions ensures the system's scalability and practicality.
The proposed SiS model achieves its best results with the ``1-2-3 (Ours)'' configuration, underscoring the advantages of multi-task learning in leveraging complete sensor data. This provides a robust, cost-effective, high-performance solution for UAV localization in complex indoor environments.

\section{CONCLUSION}
This paper introduces CiUAV, a novel and cost-effective solution for the 3D indoor localization of UAVs using CSI. Using the ESP32 IoT chip, the proposed system eliminates the need for additional onboard equipment, making it an ideal device-free approach for UAVs with strict payload constraints. The dynamic AGC compensation algorithm effectively reduces the noise from CSI. It stabilizes measurements, while the SiS multi-task model addresses the challenges of incomplete sensor data and limited training samples by optimizing sensor configurations and adapting to varying training data sizes. This approach ensures reliable localization performance in resource-constrained environments. Experiment results validate the system's performance, achieving a LMSE localization error of 0.2629 m in 3D space, highlighting its high-precision localization capability. The proposed framework provides a scalable and robust solution suitable for deployment in complex indoor environments. This work enables safe and efficient UAV operations in GPS-denied scenarios, expanding their potential for applications.


 




\vfill

\end{document}